# Single Molecule Spectral Fluctuation Originates from the Variation in Dipole Orientation Connected to Accessible Vibrational Modes

**Aranyak Sarkar,[†,§] Vinu Namboodiri,[†] Manoj Kumbhakar[†,§,]***

[†] Radiation & Photochemistry Division, Bhabha Atomic Research Centre, Mumbai 400085, India

[§] Homi Bhabha National Institute, Training School Complex, Anushaktinagar, Mumbai 400094, India

**ABSTRACT:** Fluctuation in fluorescence emission of immobilized single molecule is typically ascribed to the chromophore's intrinsic structural conformations and the influence of local environmental factors. Despite extensive research over several decades since its initial observation, a direct connection between these spectral fluctuations and the rearrangement of emission dipole orientations has remained elusive. In this study, we elucidate this fundamental molecular behavior and its underlying mechanisms by employing unique single-molecule multidimensional tracking to simultaneously monitor both the emission spectrum and the three-dimensional dipole orientation of individual fluorophore. For the first time, we present compelling evidence demonstrating a correlation between spectral fluctuations and dipolar rearrangements at room temperature. Our observations reveal that variations in the radiative relaxation probabilities among different vibronic emission bands, coupled with the interaction of associated vibrational modes, drive these spectral fluctuations. We identify significant out-of-plane dipole reorientations during pronounced spectral fluctuations—commonly known as spectral jumps—which primarily arise from transitions between dominant vibrational modes. Furthermore, we emphasize the potential for constructing vibrational spectra and optical nanoscopy with vibrational specificity, leveraging the vibronic emissions from single emitters.

Advancements in optical techniques for studying single molecules (SM) at room temperature [1-3] have provided unprecedented insights into molecular interactions. A key finding is the fluctuation in SM emission spectra, attributed to the combined contributions of the molecule and its surrounding environment [4-11]. These fluctuations or more commonly known as spectral diffusion include random Gaussian variations in the spectral centroid ($\lambda_c$) around a constant mean causing spectral broadening and occurs on a very fast timescales of sub-nanoseconds. Additionally, both abrupt (spectral jumps) and continuous shifts in $\lambda_c$ have been observed, spanning timescales from milliseconds (limited by acquisition time) to several thousands of seconds until the fluorophore is photobleached. These spectral shifts are primarily due to changes in underlying electronic energy levels, driven by intrinsic factors such as inter-convertible structural conformations of the chromophore, and extrinsic factors such as interactions with the local microenvironment [9-13]. Consequently, changes in the conformation of chromophores are expected to result in shifts in spectral position and alter the orientation of molecular emission dipole. However, no correlation between spectral fluctuation and dipole orientation has been reported previously, likely due to the limited resolution of early orientation measurement techniques that intended only for in-plane angles (e.g., Sulphorhodamine 101 [9], Oxazin-1 [10]). Recent advancements in three-dimensional SM dipole orientation imaging with precision of few degrees [14-22] have prompted us to reinvestigate this aspect. At low temperatures, such spectral behavior is linked to the configurational fluctuations of surrounding molecules [8, 23-25]. Single-molecule vibrational spectroscopy at 1.8 K using fluorescence spectra of pentacene and terrylene in p-terphenyl allowed examination of individual chromophore environments [7]. Room temperature studies suggest that two vibronic bands of Rhodamine 630 and their relative intensity fluctuations contribute to spectral diffusion [26], with spectral separations close to ensemble Raman bands [27-29]. These findings indicate a connection between room temperature spectral fluctuations and chromophore intrinsic vibronic modes, offering an opportunity to re-explore the mechanistic details of spectral fluctuation through the new paradigm of SM dipole orientation and spectral measurement.

In this letter, we present the first direct evidence of a correlation between dipole orientation and spectral fluctuations, specifically spectral jumps, utilizing our novel experimental scheme. This approach enables the simultaneous tracking and correlation of 3D dipole orientation with the fluorescence spectrum of a single molecule. Our results indicate that out-of-plane changes in dipole orientation, rather than in-plane variations, primarily drive spectral jumps, thereby clarifying previously misunderstood experimental conclusions [9, 10]. By decoding vibrational information from the single-molecule fluorescence spectrum, we explore the paradigm of vibronic emissions into various thermally accessible modes and their concomitant couplings. These findings suggest that spectral jumps are manifestations of switching between dominant vibrational modes. Additionally, this study demonstrates the unique applicability of our scheme for constructing room-temperature vibrational spectra of bright fluorophores, offering a valuable gateway for retrieving chemical-specific vibrational information with the exceptional sensitivity of fluorescence for optical nanoscopy.

Spectral fluctuation of individual Rhodamine 6G (Rh6G) molecules (Fig. S4) spin coated on glass cover slip (with a thin protecting cover of poly(methyl methacrylate), PMMA [9]) were recorded using a home build spectrally resolved single molecule orientation localization microscopy (SR-SMOLM [30]) setup (Fig. 1a). Briefly, setup consists of a spectral detection channel through prism-based dispersion and one orientation-localization channel through a Vortex phase plate (VPP). Representative images of raw Vortex dipole spread functions (DSF) of four Rh6G molecules and their respective spectral dispersion for a single frame are shown in Fig. 1b. Dipole orientation parameters depicted in Fig. 1c were extracted following vectorial fitting of SM Vortex dipole spread functions outlined by Hulleman et al. [19]. Conversely, the spectral centroid ($\lambda_c$) was determined by calculating the intensity-weighted mean of the spectral intensity distribution across the camera pixels. Additionally, the vibronic signatures of individual molecules, as indicated by the double Gaussian spectral profiles, were analyzed to determine their emission maxima ($\lambda_1$ and $\lambda_2$). The spectral centroid is a standard parameter used to represent spec-tral variation over time, while the separation between the emission maxima's ($\Delta \nu$) indicate identity of associated vibronic modes. From the stack of images SM traces with 0.2 s resolution for various parameters like $\lambda_c$, $\theta$, $\phi$, $\lambda_1$, $\lambda_2$ and $\Delta \nu$ were constructed.

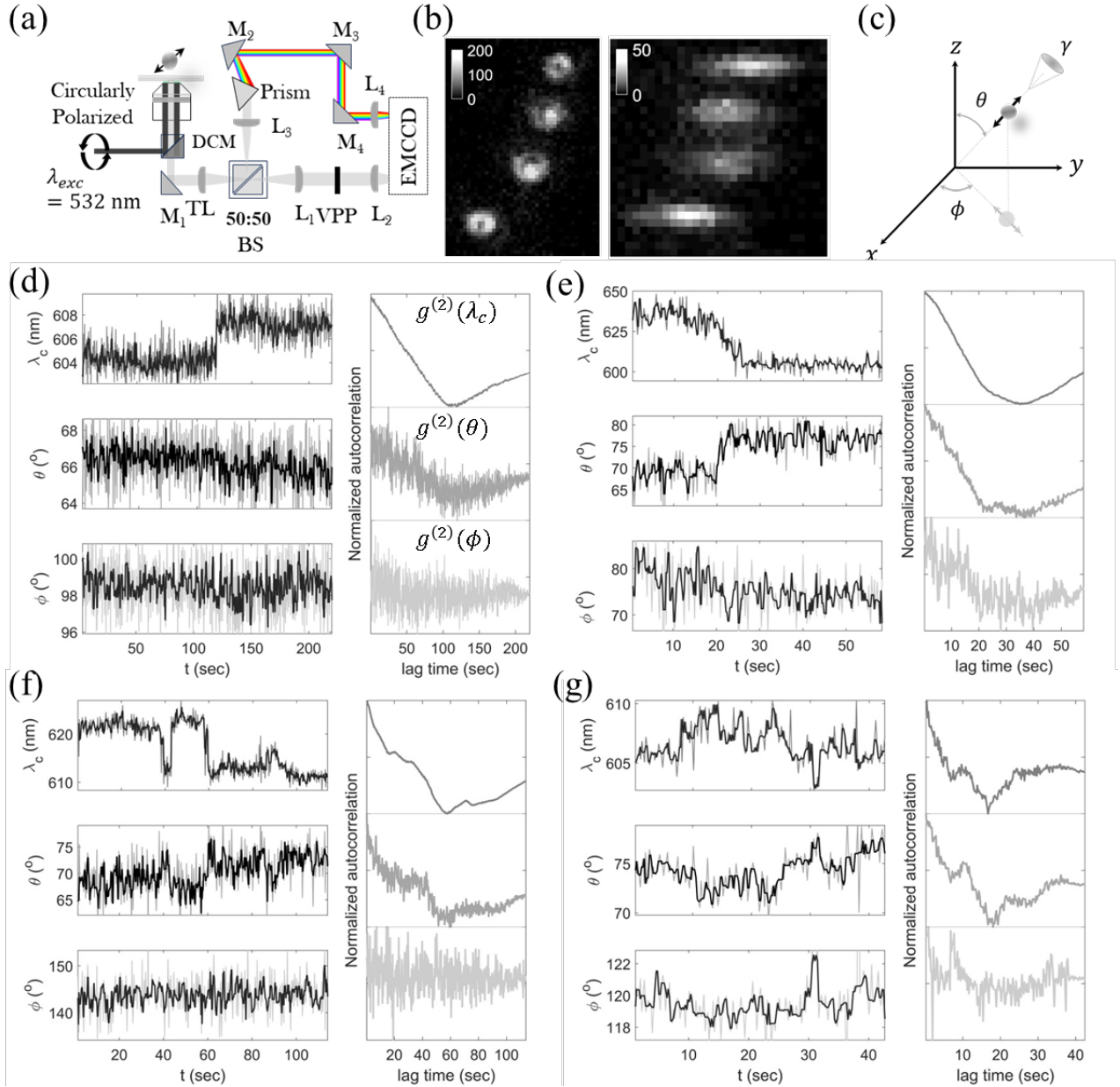

Fig. 1. (a) Optical layout of SR-SMOLM setup. DCM: Dichroic mirror, TL: Tube Lens, BS: Beam splitter, M: Mirrors, L: Lens, VPP: Vortex phase plate, EMCCD: Electron multiplying charge-coupled device. (b) Simultaneously recorded SM vortex DSF image and spectral image. (c) Coordinate system representing orientation parameters ($\theta$, $\phi$) and associated rotational constraint $\gamma$ for a given dipole emitter. (d–g) Left panels represent orientation and spectral trajectories of four SMs at different local environments, where light colors represent actual trace of parameters while the dark colors on top only act as guiding line obtained applying 3 points median filter. Right panels show respective SM auto-correlation functions.

Careful analysis of the single molecule trajectories, illustrated in Figs. 1 and S5, reveals several key aspects. We present four representative trajectories, each corresponding to a broad category of spectral diffusion: spectral jumps resembling a Heaviside step function (Fig. 1d), gradual shifts in the spectral centroid (Fig. 1e), reversible jumps (Fig. 1f), and stochastic jumps and shifts (Fig. 1g). Notably, several molecules exhibited no spectral jumps over the course of several minutes. The second-order autocorrelation of $\lambda_c$, $\theta$, and $\phi$ encodes this time evolution and depicts striking similarity. Although changes in the out-of-plane orientation angle ($\theta$) are often only 2-8° during spectral jumps (e.g. Fig. 1d), but they always coincide with changes in the spectral centroid. Occasionally, some SMs exhibit stochastic variation in $\lambda_c$ coordinated with changes in $\theta$ upon spectral jumps, along with small variations in $\phi$ (Fig. 1e). The observed close similarity between the autocorrelations of $\theta$ and $\lambda_c$ from two completely independent and simultaneous measurements confirms their correlative behavior and highlights the participation of out-of-plane dipole motions in spectral fluctuations. In line with single molecule trajectories, the overall autocorrelation function for $\lambda_c$ (Fig. S6) shows slow relaxation dynamics over seconds, possibly signifying reconfiguration of the chromophore environment leading to alterations in the spectral centroid. The origin of the observed specific anti-correlative feature between $g^{(2)}(\theta)$ and $g^{(2)}(\lambda_c)$ displayed by  all molecules (refer to Figs. S5 & S6) remains unclear. However, it is plausible that electrostatic interactions between the silica surface of the coverslip and the cationic Rh6G, which is situated near the hydrogen bond-accepting methacrylate groups of PMMA, may influence the energetics, thereby causing a shift in $\lambda_c$, in contrast to changes in polar angle. Interestingly in some occasions, a slow anti-correlative oscillatory behavior is also observed (Fig. S5), might be attributed to the back-and-forth out-of-plane motion of the chromophore coupled with the matrix.

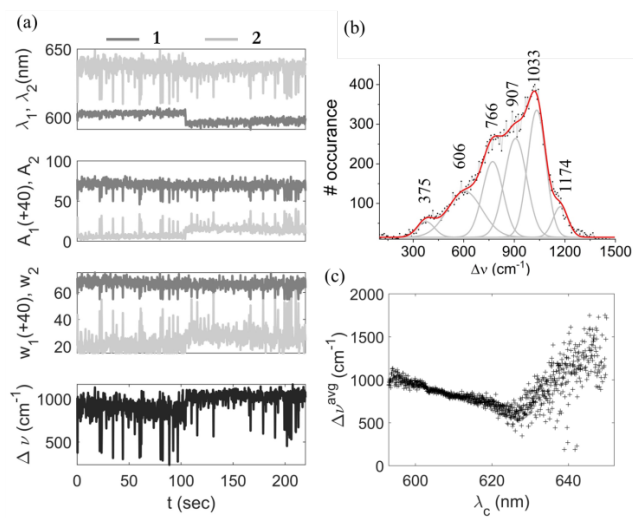

Fig. 2. (a) Trajectories of individual vibronic bands, their amplitudes, FWHMs and separation obtained from the fitting of normalized spectra in individual frames for a representative single molecule (Fig. 1d). Frame acquisition time is 0.2 s. (b) Broad vibrational spectrums constructed from more than 46000 single molecule spectral data overlaid with the closely identifiable dominant vibrational modes using multiple-peak fit routine. (c) The overall trend of the mean energy of associated vibronic bands with the spectral centroid suggests two distinct domains with respect to the spectral centroid.

To comprehend spectral behavior we looked into the variations of $\lambda_1$ and $\lambda_2$ along with their respective spectral widths (FWHM, $w$), amplitudes (integrated area, $A$) and separation ($\Delta \nu = \lambda_1^{-1} - \lambda_2^{-1}$), shown in Fig. 2a. Other than Gaussian-stochastic type fluctuation around a constant mean value, redistribution of intensity in favor of $\lambda_2$ emission band together with its hypsochromic shift (in majority molecules) and broader spectral width is clearly evident throughout the trajectory, also see Fig. S7. The combined contributions of many single molecule spectra yield a diffusion driven broad spectral distribution of $\Delta \nu$ (Fig. 2b), revealing major vibronic bands that closely resemble the reported vibrational spectra of Rh6G [31-34]. Recurrent low-energy bands observed for the majority of SMs ($\Delta \nu \sim 380$ cm$^{-1}$) likely indicate chromophore's inherent tendency to dissipate energy through electronic transitions involving torsional and bending motions [34]. The bivariate histogram between the average vibronic energy ($\Delta \nu^{avg}$) and the spectral centroid (Figs. 2c & S8) illustrates two primary variants of spectral diffusion for Rh6G. For $\lambda_c < 624$ nm, spectral shifts are primarily driven by $\lambda_1$, leading to a decrease in the spectral centroid with increasing $\Delta \nu^{avg}$. Conversely, for $\lambda_c > 624$ nm nm, the influence of $\lambda_2$ becomes dominant, resulting in an opposite trend.

In pursuit of further insights, assuming Gaussian distribution for the random fluctuations in vibronic emission maxima, we tried to ascertain identity of involved dominant higher energy vibrational bands ($\Delta \nu > 400$ cm$^{-1}$) from the average spectra of the individual molecule before and after the spectral jump(s) (Figs. 3a & S9) by filtering out the contributions from lower energy modes (shown in Fig.2a). Before the spectral jump, vibronic emission is found to exhibit a fixed separation corresponding to a particular vibronic mode (e.g. 913 cm$^{-1}$ for $t < 100$ $s$ in Fig. 3a) and transit to another mode (e.g. 1032 cm$^{-1}$ for $t > 100$ $s$) during spectral jump. Additionally, molecules exhibiting multiple spectral jumps, and thus contributions from more than two vibronic bands, were also observed (Fig. S10). The change in energy from one vibrational mode to the other is $\sim$ 100-300 cm$^{-1}$, closely matches with the available thermal energy at room temperature. Hence, it is clear that the ergodic trajectories of SM with respect to spectral fluctuations are primarily associated with time-dependent variations in thermally accessible vibronic emission bands.

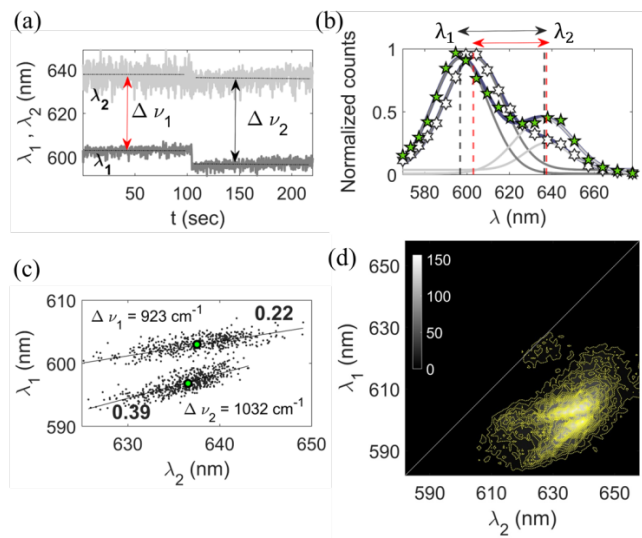

Fig. 3. (a) The filtered time trace of $\lambda_1$ and $\lambda_2$, shown in Fig. 2a. The black dotted line on top represents the fitted values of these components averaged over the respective time range. (b) Normalized time-averaged fitted spectra with two emission maxima before and after the spectral jump at approximately $t$=100 seconds. (c) 2D plot of vibronic band maxima of the same single molecule random telegraph shows two distinct populations. Two circles at the center of the distributions represent respective average values before and after the jump. The lines represent linear fit of the two vibronic populations. (d) Overall 2D plot of vibronic band maxima of all single molecules.

The 2D plot of $\lambda_1$ & $\lambda_2$ (Fig. 3c) for the same single molecule underscore the variation in coupling among various accessible vibronic modes. Most importantly higher energy vibrational modes are relatively less coupled (slope of 0.39 for 1032 cm$^{-1}$) than lower energy vibrational modes (slope of 0.22 for 923 cm$^{-1}$). Even for a particular vibrational mode we observed variation in slope value from molecule to molecule (Fig. S11), suggests inhomogeneity in local environment. The overall 2D plot of vibronic band maxima (Fig. 3d) from all single molecules clearly shows two separate vibronic populations with dissimilar vibronic relaxation characteristics. It reflects relatively greater coupling among lower energy vibrational modes than with the higher energy vibrational modes during $S_1 \rightarrow S_0$ transition. A small population of nearly close vibronic band positions at ~625 nm is also noted, highlighting the appearance of low frequency vibrational modes at around 100-150 cm$^{-1}$, generally assigned to modes involving atoms in hydrogen bonds [35, 36].

Retrieval of all the involved vibrational modes accessed by Rh6G in the finger-print region was done employing a simple 50 point averaging tool (described in Fig. S12), and applied to all single molecule $\lambda_1,\lambda_2$ traces to construct their individual vibrational spectrum (Figs. S10, S11) [37]. The overall vibrational spectrum is shown in Fig. 4a, which matches with the vibrational modes of Rh6G in the finger-print region recorded with ensemble ground state Raman spectroscopy (Table S1) [31-34]. Additionally, observed 913 cm$^{-1}$ band (Fig. 4a) corroborates the coupling of Rh6G excited state ($S_1$) in the selection of active vibronic modes, which is absent in ground state Raman scattering spectrum [31]. Many spectroscopic methods explore only a small portion of the potential energy surface and thus contain very limited information. Here too the vibrational information contained in the emission spectrum is limited to the states accessible with the excitation laser used. Unlike narrowing of emission lines at cryogenic temperatures to identify [38, 39] and record high-resolution vibrational spectra [40], here measurements are done at room temperature only to characterize chemical signature.

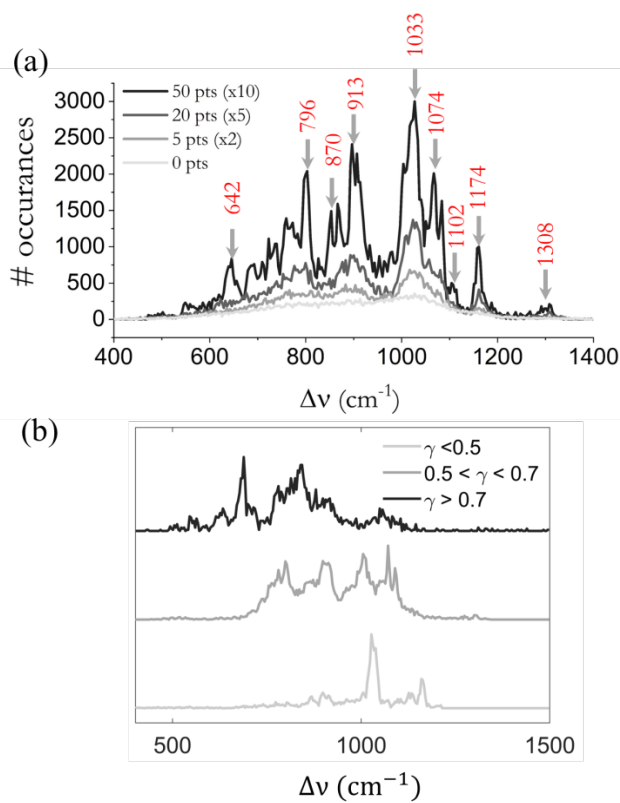

Fig. 4. (a) Combined vibrational spectra obtained from individual single molecule spectral trajectories with 50 points median filter [37]. (b) Segregated vibrational spectrum in different γ ranges.

Further, a quick segregation of vibrational spectrum based on γ values (Fig. 4b) display dominance of low energy vibrational modes for molecules experiencing highly restricted environment, while relatively less constraint molecules exhibit dominance of higher energy vibrational modes. Interestingly, higher energy vibrational bands also display relatively narrow width compared to low energy vibrational bands. This is indicative of greater coupling of low frequency vibrational modes with chromophore's fluctuating local hydrogen-bonded network involving PMMA and under suppressed librational flexibility within the nanocavity of silica and protecting polymer layer [41, 42]. Both these factors are known to influence vibrational coupling. This corroborates our earlier observation of direct connection between dipole orientation fluctuation and the observed vibronic band populations and suggests possibility to tune vibrational modes through confinement modulation.

At room temperature a chromophore in its nanosecond long excited state explores many local configurations with its surrounding and vibrational states. However, identification of vibronic modes during its transition to the ground state is possible by means of fast single molecule spectral imaging. Of course resolution of such room temperature vibrational spectrum is compromised with respect to spectral diffusion experiments at very low temperatures but on the contrary offers an excellent opportunity to access chemical specific information along with the huge sensitivity of fluorescence suitable to adopt even for biological samples. Attaining better than 10 nm localization precision with bright fluorophore is trivial in current perspective and hence optical nanoscopy (super-resolution fluorescence microscopy) with chemical specific information is not far from reality [43].

In conclusion, present endeavor of single molecule multi-parameter tracking demonstrates definite correlation between dipole orientation and spectral wandering linked to intrinsic vibrational modes supported by the local environment through vibronic emission. Such multi-parameter dynamics can also be employed to study molecular interactions. Tracking intrinsic vibrations will not only give characteristic attributes of the carefully chosen dye but will also give information on the structure and dynamics of its local environment by revealing polarity, mobility, reorientation constrain, couplings with its surroundings, etc. It is also possible to construct images with nanoscopic resolution and chemical specific information by employing suitable fluorophore from the already available vast color pallet.

## ASSOCIATED CONTENT

### Supporting Information

Additional figures for spectral fluctuations, analysis methodology, measurement accuracy & precession, sample preparation (PDF).

## AUTHOR INFORMATION

### Corresponding Author

MK: manojk@barc.gov.in

### Notes

The authors declare no competing financial interests.

## ACKNOWLEDGMENT

This work is funded and supported by BARC (DAE), India.

## REFERENCES

[1] W. E. Moerner, and L. Kador, Optical Detection and Spectroscopy of Single Molecules in A Solid, Phys. ReV. Lett. **62**, 2535 (1989).

[2] M. Orrit, and J. Bernard, Single Pentacene Molecules Detected by Fluorescence Excitation in a p-Terphenyl Crystal, Phys. ReV. Lett. **65**, 2716 (1990).

[3] E. B. Shera, N. K. Seitzinger, L. M. Davis, R. A. Keller, and S. A. Soper, Detection of Single Fluorescent Molecules, Chem. Phys. Lett. **174**, 553 (1990).

[4] W. E. Moerner, A Dozen Years of Single-Molecule Spectroscopy in Physics, Chemistry, and Biophysics, J. Phys. Chem. B **106**, 910 (2002).

[5]     W.-T. Yip, D. Hu, J. Yu, D. A. V. Bout, and P. F. Barbara, Classifying the Photophysical Dynamics of Single- and Multiple-Chromophoric Molecules by Single Molecule Spectroscopy, J. Phys. Chem. A **102**, 7564 (1998).

[6]     X. S. Xie, and R. C. Dunn, Probing Single Molecule Dynamics, Science **265** 361 (1994).

[7]     A. B. Myers, P. Tchénio, M. Z. Zgierski, and W. E. Moerner, Vibronic Spectroscopy of Individual Molecules in Solids, J. Phys. Chem. **98**, 10377 (1994).

[8]     R. Brown, and M. Orrit, in *Single-Molecule Optical Detection, Imaging and Spectroscopy*, edited by T. Basché *et al.* (VCH Verlagsgesellschaft GmbH, Weinheim, 1997).

[9]     H. P. Lu, and X. S. Xie, Single-molecule Spectral Fluctuations at Room Temperature, Nature **385** 143 (1997).

[10]    C. Jung, C. Hellriegel, B. Platschek, D. W. hrle, T. Bein, J. Michaelis, and C. Bra¨uchle, Simultaneous Measurement of Orientational and Spectral Dynamics of Single Molecules in Nanostructured Host-Guest Materials, J. Am. Chem. Soc. **129**, 5570 (2007).

[11]    G. Sallen, A. Tribu, T. Aichele, R. Andre´, L. Besombes, C. Bougerol, M. Richard, S. Tatarenko, K. Kheng, and J.-P. Poizat, Subnanosecond Spectral Diffusion Measurement Using Photon Correlation, Nat. Photonics **4** 696 (2010).

[12]    M. Brecht, H. Studier, V. Radics, J. B. Nieder, and R. Bittl, Spectral Diffusion Induced by Proton Dynamics in Pigment-Protein Complexes, J. Am. Chem. Soc. **130**, 17487 (2008).

[13]    A. M. v. Oijen, M. Ketelaars, J. K. hler, T. J. Aartsma, J. Schmidt, Spectroscopy of Individual Light-Harvesting 2 Complexes of Rhodpseudomonas acidophila: Diagonal Disorder, Intercomplex Heterogeneity, Spectral Diffusion, and Energy Transfer in the B800 Band, Biophys. J. **78** 1570 (2000 ).

[14]    D. Patra, I. Gregor, and J. Enderlein, Image Analysis of Defocused Single-Molecule Images for Three-Dimensional Molecule Orientation Studies, J. Phys. Chem. A **108**, 6836 (2004).

[15]    C. V. Rimoli, C. A. Valades-Cruz, V. Curcio, M. Mavrakis, and S. Brasselet, 4polar-STORM Polarized Super-resolution Imaging of Actin Filament Organization in Cells, Nat. Comm. **13**, 301 (2022).

[16]    C. A. V. Cruz, H. A. Shaban, A. Kress, N. Bertaux, S. Monneret, M. Mavrakis, J. Savatier, and S. Brasselet, Quantitative Nanoscale Imaging of Orientational Order in Biological Filaments by Polarized Superresolution Microscopy, Proc. Natl. Acad. Sci. **113**, E820 (2016).

[17]    O. Zhang, W. Zhou, J. Lu, T. Wu, and M. D. Lew, Resolving the Three-Dimensional Rotational and Translational Dynamics of Single Molecules Using Radially and Azimuthally Polarized Fluorescence, Nano Lett. **22**, 1024 (2022).

[18]    O. Zhang, Z. Guo, Y. He, T. Wu, M. D. Vahey, and M. D. Lew, Six-Dimensional Single-Molecule Imaging with Isotropic Resolution Using a Multi-View Reflector Microscope, Nat. Photonics **17**, 179 (2023).

[19]    C. N. Hulleman, R. Ø. Thorsen, E. Kim, C. Dekker, S. Stallinga, and B. Rieger, Simultaneous Orientation and 3D Localization Microscopy with a Vortex Point Spread Function, Nat. Comm. **12**, 5934 (2021).

[20]    T. Ding, and M. D. Lew, Single-Molecule Localization Microscopy of 3D Orientation and Anisotropic Wobble Using a Polarized Vortex Point Spread Function, J. Phys. Chem. B **125**, 12718 (2021).

[21]    A. Sarkar, V. Namboodiri, and M. Kumbhakar, Single-Molecule Orientation Imaging Reveals Two Distinct Binding Configurations on Amyloid Fibrils, J. Phys. Chem. Lett. **14**, 4990 (2023).

[22]    M. P. Backlund, M. D. Lew, A. S. Backer, S. J. Sahl, and W. E. Moerner, The Role of Molecular Dipole Orientation in Single-Molecule Fluorescence Microscopy and Implications for Super-Resolution Imaging, ChemPhysChem **15**, 587 (2014).

[23]    T. Basche´, and W. E. Moerner, Optical Modification of a Single Impurity Molecule in a Solid, Nature **355**, 335 (1992).

[24]    W. P. Ambrose, T. Basché, and W. E. Moerner, Detection and Spectroscopy of Single Pentacene Molecules in a p-Terphenyl Crystal by Means of Fluorescence Excitation, J. Chem. Phys. **95**, 7150 (1991).

[25]    A. Zumbusch, L. Fleury, R. Brown, J. Bernard, and M. Orrit, Probing Individual Two-Level Systems in a Polymer by Correlation of Single Molecule Fluorescence, Phys. Rev. Lett. **70**, 3584 (1993).

[26]    A. J. Meixner, and M. A. Weber, Single Molecule Spectral Dynamics at Room Temperature, J. Luminescence **86** 181 (2000).

[27]    F. Stracke, C. Blum, S. Becker, K. Müllen, and A. J. Meixner, Intrinsic Conformer Jumps Observed by Single Molecule Spectroscopy in Real Time, Chem. Phys. Lett. **325** 196 (2000).

[28]   C. Blum, F. Stracke, S. Becker, K. Müllen, and A. J. Meixner, Discrimination and Interpretation of Spectral Phenomena by Room-Temperature Single-Molecule Spectroscopy, J. Phys. Chem. A **105**, 6983 (2001).

[29]   F. Stracke, C. Blum, S. Becker, K. Müllen, and A. J. Meixner, Two and Multilevel Spectral Switching of Single Molecules in Polystyrene at Room Temperature, Chem. Phys. **300** 153 (2004).

[30]   A. Sarkar, J. B. Mitra, V. K. Sharma, V. Namboodiri, and M. Kumbhakar, Spectrally Resolved Single Molecule Orientation Imaging Reveals Direct Correspondence between Polarity and Order Experienced by Nile Red in Supported Lipid Bilayer Membrane, bioRxiv 2024.06.21.600028; doi: https://doi.org/10.1101/2024.06.21.600028 (2024).

[31]   L. Jensen, and G. C. Schatz, Resonance Raman Scattering of Rhodamine 6G as Calculated Using Time-Dependent Density Functional Theory, J. Phys. Chem. A **110** 5973 (2006).

[32]   J. Guthmuller, and B. t. Champagne, Resonance Raman Scattering of Rhodamine 6G as Calculated by Time-Dependent Density Functional Theory: Vibronic and Solvent Effects, J. Phys. Chem. A **112**, 3215 (2008).

[33]   H. Watanabe, N. Hayazawa, Y. Inouye, and S. Kawata, DFT Vibrational Calculations of Rhodamine 6G Adsorbed on Silver: Analysis of Tip-Enhanced Raman Spectroscopy, J. Phys. Chem. B **109**, 5012 (2005).

[34]   S. Shim, C. M. Stuart, and R. A. Mathies, Resonance Raman Cross-Sections and Vibronic Analysis of Rhodamine 6G from Broadband Stimulated Raman Spectroscopy, ChemPhysChem **9**, 697 (2008).

[35]   S. E. M. Colaianni, and O. F. Nielsen, Low-Frequency Raman Spectroscopy, J. Mol. Struc. **347** 267 (1995).

[36]   S. Kim, X. Wang, J. Jang, K. Eom, S. L. Clegg, G.-S. Park, and D. D. Tommaso, Hydrogen-Bond Structure and Low-Frequency Dynamics of Electrolyte Solutions: Hydration Numbers from ab Initio Water Reorientation Dynamics and Dielectric Relaxation Spectroscopy, ChemPhysChem **21**, 2334 (2020).

[37]   M. Weber, H. v. d. Emde, M. Leutenegger, P. Gunkel, S. Sambandan, T. A. Khan, J. Keller-Findeisen, V. C. Cordes, and S. W. Hell, MINSTED nanoscopy enters the Ångström localization range, Nature Biotechnology (2022).

[38]   T. Liu, B. Carles, C. Elias, C. Tonnelé, D. Medina-Lopez, A. Narita, Y. Chassagneux, C. Voisin, D. Beljonne, S. Campidelli, L. Rondin, and J.-S. Lauret, Vibronic Fingerprints in the Luminescence of Graphene Quantum Dots at Cryogenic Temperature, J. Chem. Phys. **156**, 104302 (2022).

[39]   A. Kiraz, M. Ehrl, C. B. uchle, and A. Zumbusch, Low Temperature Single Molecule Spectroscopy Using Vibronic Excitation and Dispersed Fluorescence Detection, J. Chem. Phys. **118**, 10821 (2004).

[40]   J. Zirkelbach, M. Mirzaei, I. Deperasińska, B. Kozankiewicz, B. Gurlek, A. Shkarin, T. Utikal, S. Götzinger, and V. Sandoghdar, High-resolution Vibronic Spectroscopy of a Single Molecule Embedded in a Crystal, J. Chem. Phys. **156**, 104301 (2022).

[41]   C.-C. Yu, K.-Y. Chiang, M. Okuno, T. Seki, T. Ohto, X. Yu, V. Korepanov, H.-o. Hamaguchi, M. Bonn, J. Hunger, and Y. Nagata, Vibrational Couplings and Energy Transfer Pathways of Water's Bending Mode, Nat. Comm. **11**, 5977 (2020).

[42]   P. K. Verma, A. Kundu, M. S. Puretz, C. Dhoonmoon, O. S. Chegwidden, C. H. Londergan, and M. Cho, The Bend+Libration Combination Band Is an Intrinsic, Collective, and Strongly Solute-Dependent Reporter on the Hydrogen Bonding Network of Liquid Water, J. Phys. Chem. B **122**, 2587 (2018).

[43]   A preliminary proof of principle demonstration for single molecule spectral and vibrational imaging along with their 3D dipole orientation and nanoscopic localization is shown in Fig. S13 and discussed in the context of vibrational nanoscopy.

# Supplementary Information

# Single Molecule Spectral Fluctuation Originates from the Variation in Dipole Orientation Connected to Accessible Vibrational Modes

Aranyak Sarkar,[†,§] Vinu Namboodiri,[†] Manoj Kumbhakar[†,§,*]

[†] Radiation & Photochemistry Division, Bhabha Atomic Research Centre, Mumbai 400085, India

[§] Homi Bhabha National Institute, Training School Complex, Anushaktinagar, Mumbai 400094, India

**1. Materials**: Dye Rh6G and poly(methyl methacrylate) (PMMA, MW 3,50,000) were purchased from Sigma Aldrich. Spectroscopy grade $CHCl_3$ and HPLC grade water were obtained from Merck Life Science Pvt. Ltd. All chemicals were used as received without further purification. Cover slips (Ted Pella Inc., No. 1 thick) were cleaned as described earlier [1, 2]. Dilute dye solutions (~50 pM) were prepared in $CHCl_3$ by successive dilution.

**2. SR-SMOLM Setup**: The experiments were performed on a home built wide-field epifluorescence microscope (Axio-observer D1, Carl Zeiss GmbH) based on circularly polarized TIR excitation, as described in Ref. [1]. The excitation beam (532 nm, CW, ~1.2 kW/cm$^2$; Shanghai Dream Lasers Technology Co Ltd.) was focused on the back focal plane of the objective (100x NA1.46 oil immersion, Carl Zeiss GmbH). The fluorescence signal was collected through the dichroic (FF538-FDi01, Semrock) & filters (532 Notch filter, Semrock, 561 nm Long Pass, Semrock and a 650 nm Short Pass, Semrock). Briefly, setup (shown in Fig. 1a) consists of a spectral detection channel through prism-based dispersion and one orientation-localization channel through a Vortex phase plate (VPP, V-593-10-1, Vortex Photonics), positioned at the Fourier plane (FP) of the 4f-imaging system [3], resulting in a magnification factor of 2x and achieving a pixel size of 80 nm. Beam-splitter (BS) splits fluorescence signal after passing through the dichroic mirror (DCM) and imaged onto EMCCD camera (DU971N-UVB, Andor) through mirrors (M) and lenses (L) along the depicted two detection paths. Setup calibration and image analysis are described in Ref. [1]. Image acquisition involved recording image stacks with a 200 ms exposure time, comprising around 500 to 2,000 sequential frames.

## 3. Orientation Imaging

For the extraction of orientation parameters, we have adopted the methodology elucidated by Hulleman et al. [3] and discussed with respect to our SR-SMOLM setup in Ref. [1]; and briefly mentioned below.

### 3a. DSF Model

A single molecule dipole spread function (DSF) recorded in camera is dependent both on 3D position vector $\boldsymbol{r} \equiv (x, y, z)$ and 3D orientation parameters $\Theta \equiv (\theta, \phi, \gamma)$ as depicted below.

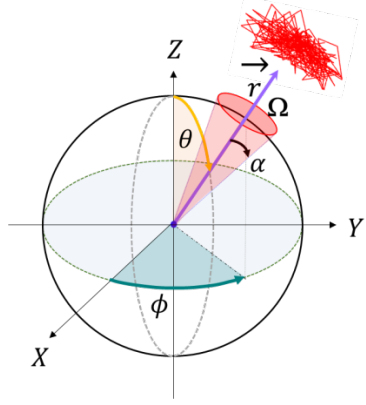

Typical frame rate of single molecule imaging lies between 10-500 ms and within the integration time of each camera frame, the molecule can access different orientation states. Thus recorded image is always a time-averaged image integrated over accessible solid angle $\Omega$. In terms of half-cone angle $\alpha$, this can written as

$$\Omega = 2\pi(1 - cos\alpha)$$

And this is related to the rotational constraint $\gamma$ through

$$\gamma = 1 - \frac{3\Omega}{4\pi} + \frac{\Omega^2}{8\pi^2}$$

referring to the fact that

$$\alpha = 0^o, \Omega = 0, \ \gamma = 1 \rightarrow \text{fixed dipole}$$

$$\alpha = 180^o, \Omega = 4\pi, \ \gamma = 0 \rightarrow \text{free dipole}$$

Therefore, DSF model $\rho(\boldsymbol{r}|\Theta)$ can assumed to be composed of a weighted sum of the contribution from a free dipole $\rho_{free}(\boldsymbol{r})$ and a rotationally immobile or fixed dipole $\rho_{fixed}(\boldsymbol{r}|\Theta)$, i.e.,

$$\rho(\boldsymbol{r}|\Theta) = N\left[\frac{(1-\gamma)}{3}\rho_{free}(\boldsymbol{r}) + \frac{\gamma}{3}\rho_{fixed}(\boldsymbol{r}|\Theta)\right] + \frac{\bar{b}}{n^2}$$

Where the pixelated intensity distribution of the single molecule DSF is assumed to be confined in $n \times n$ camera pixels and $\bar{b}$ being the background photons per pixel. $N$ is the integrated photons counts or brightness of the molecule given by

$$N = \int\limits_{x_o-n/2}^{x_o+n/2} \int\limits_{y_o-n/2}^{y_o+n/2} dxdy \ \rho(\boldsymbol{r} - \boldsymbol{r_0}|\Theta)$$

Here $\boldsymbol{r_0} \equiv (x_0, y_0)$ is center of the DSF. Now DSF of a fixed emitter can be expressed as

$$\rho_{fixed}(\boldsymbol{r}|\Theta) = \sum_{i,j \in x,y,z} A_{ij}(\boldsymbol{r})\hat{\mu}_i(\Theta)\hat{\mu}_j(\Theta)$$

where $\hat{\mu}_i$ is the unit vector along dipole direction i.e. $\hat{\mu}_i = \frac{\mu}{|\mu|} \equiv (sin\theta cos\phi, sin\theta sin\phi, cos\theta)$. Average DSF of an isotropic free dipole emitter is

$$\rho_{free}(\boldsymbol{r}) = \frac{1}{3} \sum_{i \in x,y,z} A_{ii}(\boldsymbol{r})$$

With $A_{ij}(\boldsymbol{r}) = \sum_{\boldsymbol{k} \in x,y}(E_{ki}^{img}(\boldsymbol{r}))^* E_{kj}^{img}(\boldsymbol{r})$. $E_{ki}^{img}$ are the components of electric field emitted by the dipole in image plane.

Following the fact that these field components present in the image plane ($E^{img}$) can be written in terms the Fourier transformation of the field in the back-focal plane ($E^{BFP}$) of the objective

$$E_{kj}^{img}(\boldsymbol{r}) = \iint_{-\infty}^{\infty} E_{kj}^{BFP}(\hat{\boldsymbol{r}})\Psi(\hat{\boldsymbol{r}})e^{i\boldsymbol{k}.\hat{\boldsymbol{r}}}d\hat{x}d\hat{y}$$

Now considering that, there is no amplitude modulation involved, we can take it as a pure phase-mask and can be expressed as

$$\Psi(\hat{\boldsymbol{r}}) = e^{i\psi(\hat{\boldsymbol{r}})} = \varphi(\hat{\boldsymbol{r}}) + P(\hat{\boldsymbol{r}})$$

Where $\varphi(\hat{\boldsymbol{r}})$ corresponds to the phase term due to the usage of any phase modulation optics in the detection path. In our case, this will be the zone function of vortex phase mask $(\hat{\boldsymbol{r}}) = \frac{\beta}{2\pi}$ ; $\beta = tan^{-1}\left(\frac{\hat{x}}{\hat{y}}\right)$

$P(\hat{\boldsymbol{r}})$ corresponds to the aberration term. Mathematically, this can be expressed as a linear sum of RMS (root mean square) normalized Zernike polynomial $Z_n^m(\hat{\boldsymbol{r}})$:

$$P(\hat{\boldsymbol{r}}) = \sum_{m,n} A_n^m Z_n^m(\hat{\boldsymbol{r}})$$

We adopt the methodology proposed by Rieger et al., considering the dependence on the position in the field of view (FOV). In other words, we treat the Zernike coefficients as functions of the field coordinates $(x, y)$. These functions, denoted as $A_n^m(x, y)$, are determined through a calibration procedure. We generate a through-focus image stack of beads randomly distributed across the FOV. For each bead, we retrieve the Zernike coefficients. According to Nodal Aberration Theory (NAT), the aberration coefficients $A_n^m(x, y)$ can be adequately described by low-order Taylor series in $x$ and $y$

$$A_n^m(x, y) = \sum_{jk} \xi_{nmjk} x^j y^k$$

The coefficients $nmjk$ of these Taylor series for different positions are interrelated, which proves advantageous as it reduces the number of parameters to be determined experimentally. The NAT-model is fitted to the measured $A_n^m$ at the positions of the beads using a simple least-squares fit. Through this calibration procedure, the Zernike coefficients can be effectively interpolated over the entire imaging field. Here we have considered Zernike modes with $n + |jm| \le 6$, encompassing primary and secondary astigmatism, coma, spherical aberration, and trefoil. We utilize polynomials in the field coordinates up to order $6 - n$ in the NAT description of the field dependence of the relevant Zernike modes.

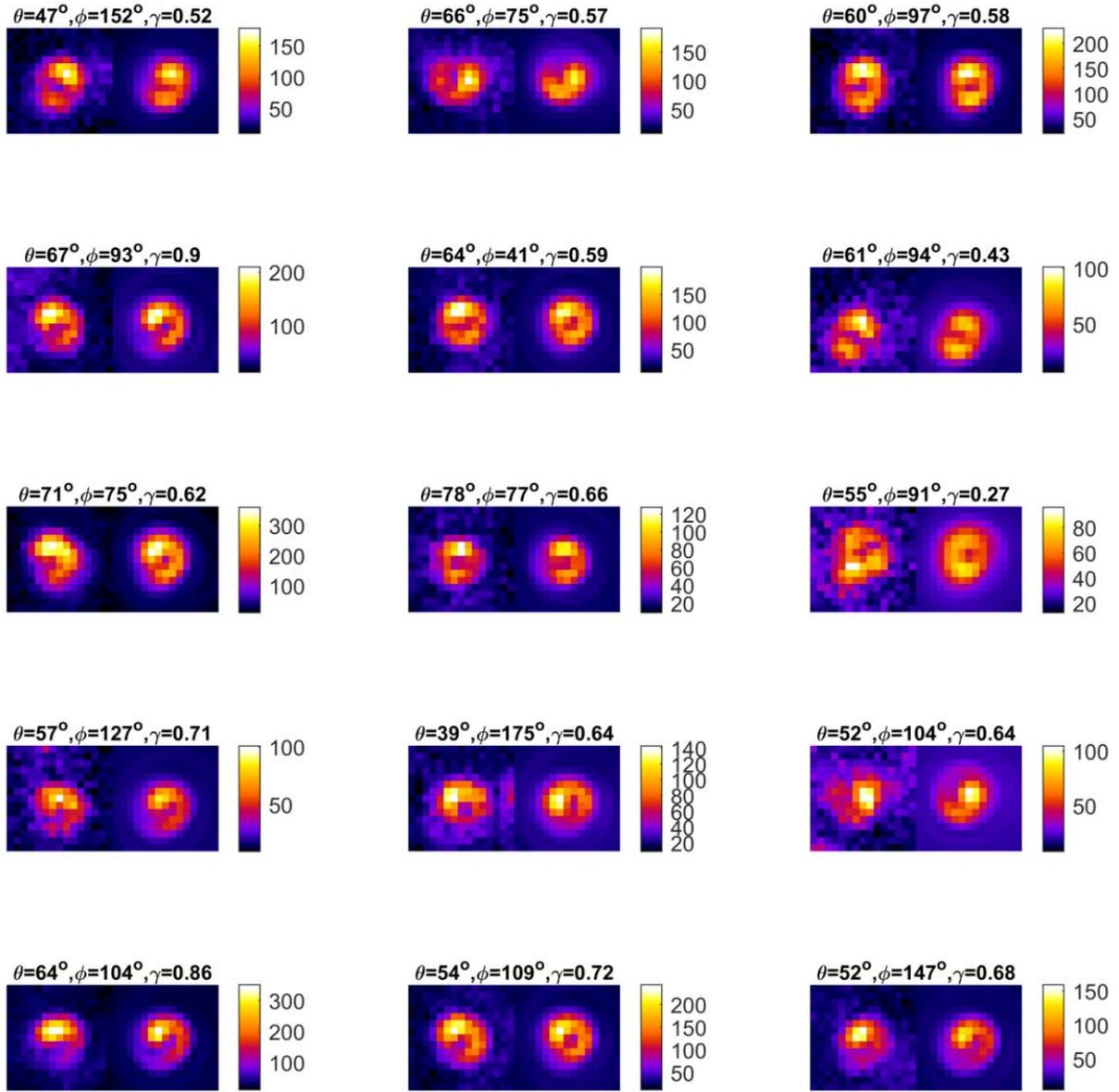

**Fig. S1.** Simulation to estimate accuracy and precisions of Orientation parameters

## 3b. DSF fitting

Intensity profile recorded by the camera $I(x_0, y_0)$ is fitted by above DSF model $\rho(\boldsymbol{r}|\Theta)$ using maximum likelihood estimation following an optimization problem

$$\widehat{\Theta} = \underset{\Theta}{\arg\max} \sum_{(x,y)\in\mathfrak{D}} \{I(x_0, y_0)\ln(\rho(\boldsymbol{r}|\Theta)) - \rho(\boldsymbol{r}|\Theta)\}$$

Few of the fitted DSF profiles along with estimated orientation parameters are shown in Fig. S1.

Analysis presented in this manuscript is based on more than 46000 recoded snaps of single molecule DSFs containing more than 400 time traces of individual bright molecules with photon counts more than 1000. Histograms of molecular brightness and background per pixel are shown in Fig. S2. Median of the brightness is ~ $6.57 \times 10^3$ photons with ~8 background photons per pixel. These values were used for simulation to study accuracy and precision of our estimation.

## 4. Spectral fitting

All the single molecule normalized spectra are fitted with double Gaussian function as given below

$$y = y_0 + \frac{1}{\sqrt{\pi/2}} \sum_{i=1}^{2} \frac{A_i}{w_i} e^{-\frac{2(\lambda - \lambda_i)^2}{w_i^2}}$$

Here $y_0$ is a constant offset, $A_i$ are the area enclosed by the Gaussian profiles centered at $\left(\lambda_i, y_{c_i}\right)$ and $w_i$ are the respective widths where

$$y_{c_i} = y_0 + \frac{A_i}{w_i \sqrt{\pi/2}}$$

The intensity-weighted centroid of the spectral intensity distribution across the camera pixels were calculated following $\lambda_c = \frac{\sum_i \lambda_i S_i}{\sum_i S_i}$, where $\lambda_i$ and $S_i$ are the wavelengths and signal of i$^{th}$ pixel.

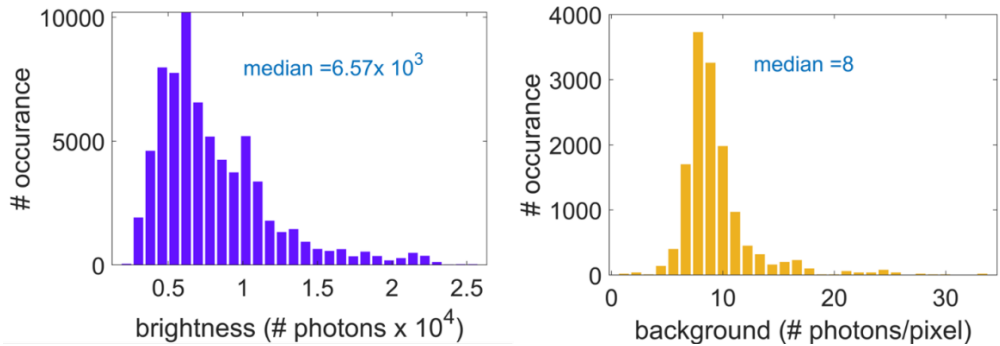

**Fig. S2. .** Bright Rh6G molecules (~ 46000) with photon count more than 1000 are only considered for the analysis presented here. Median of the brightness is ~ $6.57 \times 10^3$ photons with ~8 background photons per pixel. These values were used for simulation to study accuracy and precision of our estimation. Results are shown below.

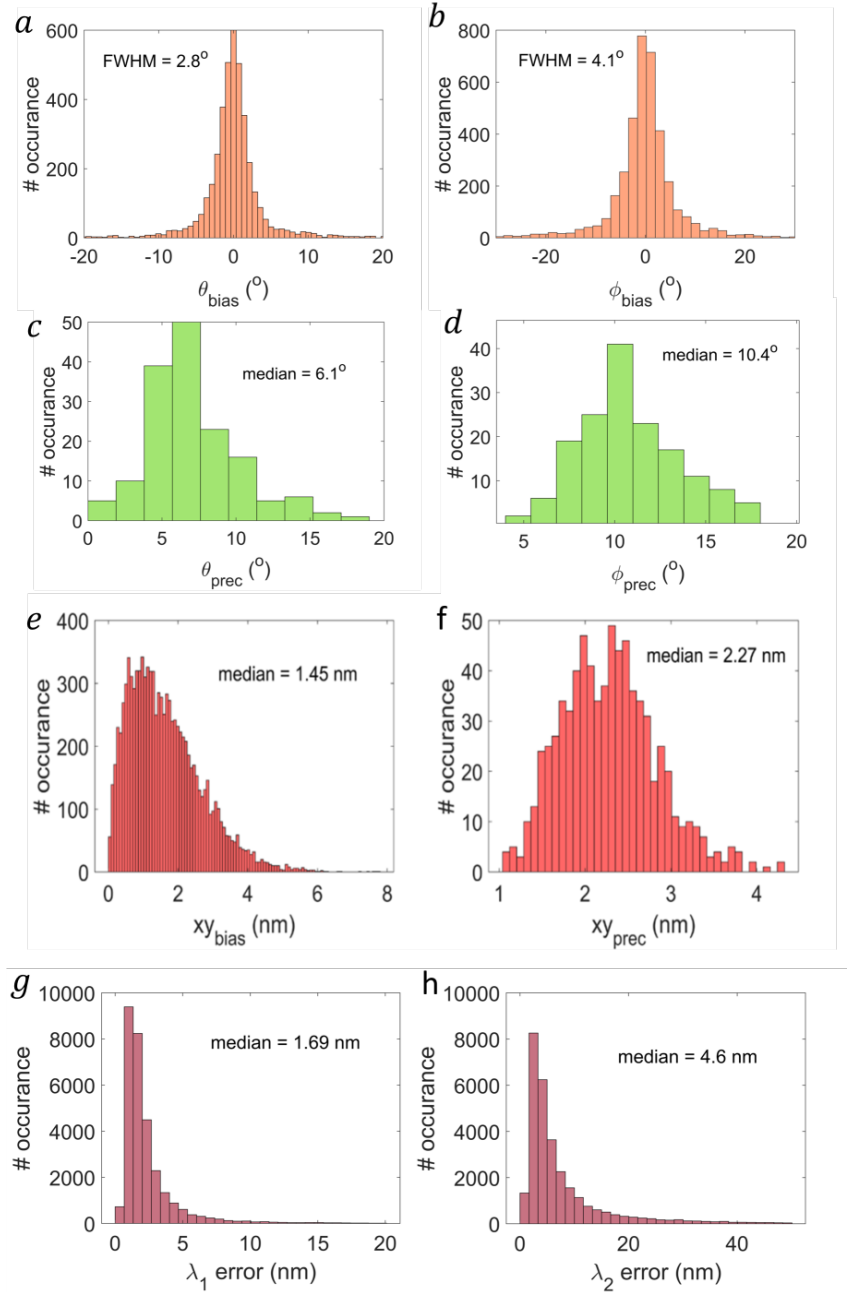

**Fig. S3.** For estimating accuracy and precision of estimated parameters, 10000 Vortex DSFs were simulated with mean brightness and background per pixel following the distributions depicted in Fig. S2. For each individual DSF, orientation and localization parameters $(\theta, \phi, x, y)$ were estimated using the method described in Ref. [1, 3], via Zernike polynomial based phase inclusion in the imaging model. Average bias (accuracy) of parameters was estimated by calculating mean difference between actual and estimated value over full $\theta$ range $(0, \pi/2)$ and $\phi$ range $(0, \pi)$. Similarly, $x$ and $y$ range are chosen to cover full field of view. Results are plotted in the bias histogram (a), (c) and (e). For estimating precision, average standard deviation of each estimated parameters was calculated in any specific bin and corresponding histograms were plotted (b), (d) and (f). Figure (g) and (h) shows histogram of the fitting error of components $\lambda_1$ and $\lambda_2$ estimated using Jacobian matrix approach.

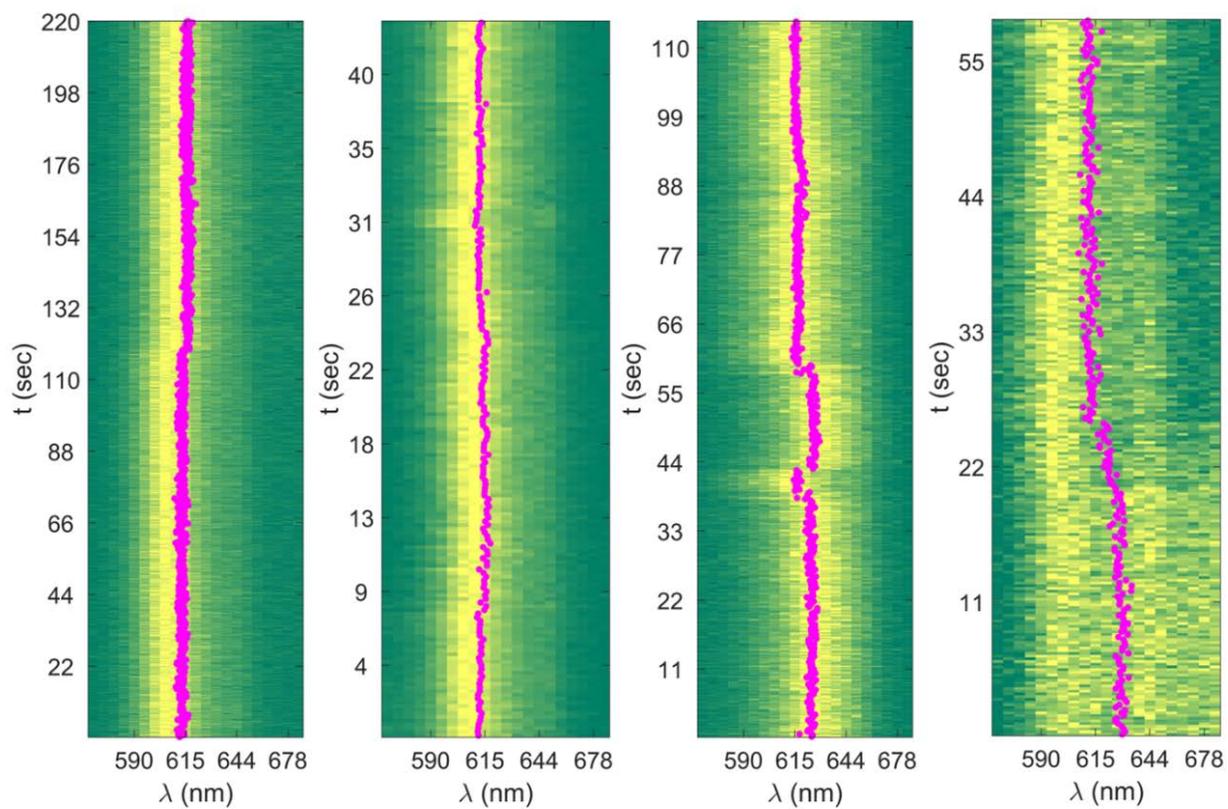

**Fig. S4.** Figure represents spectral trajectories of four SMs shown in Fig. 1. Magenta dot on each individual spectrum shows the position of the emission centroid.

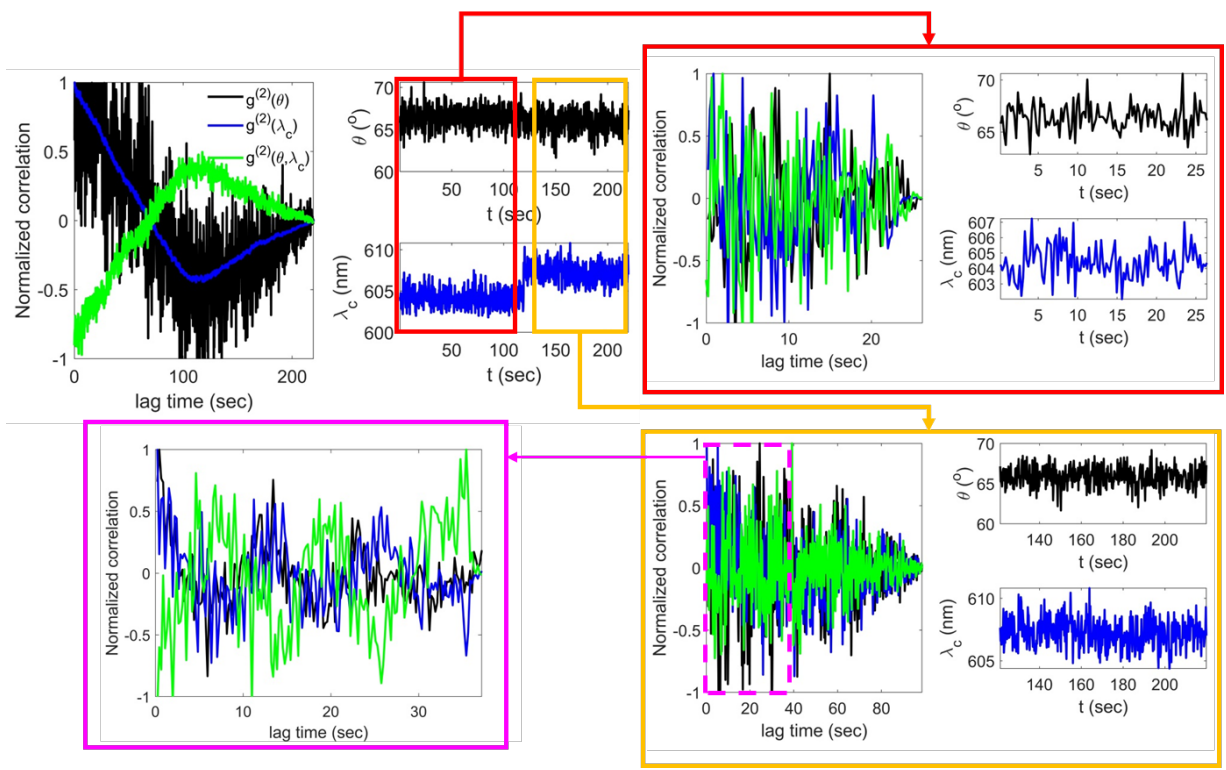

**Fig. S5.** A representative single molecule autocorrelation of polar angle $\theta$, spectral centroid $\lambda_c$ and their cross-correlation are highlighted in the top left panel along with their trajectories. This particular SM exhibits a spectral jump at $t \sim 120$ s. Top right and bottom right panel shows their time segregated trajectories and correlations before and after the jump, respectively, excluding the jump segment. Notably, both the segments shows fast stochastic fluctuation around a mean value along with a slow oscillatory behaviour in the correlations (bottom left panel) arising possibly from the back and forth movement of the molecule around the trap centre.

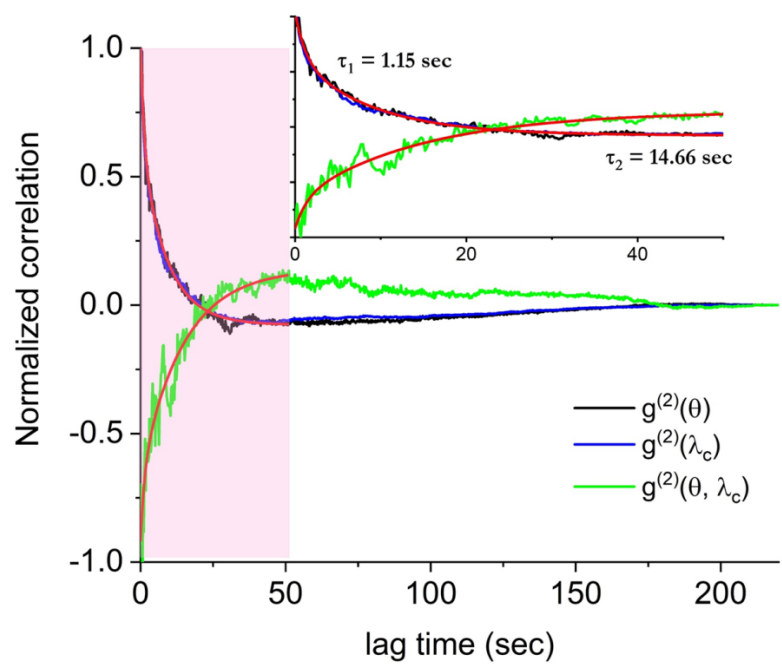

**Fig. S6.** Overall autocorrelation curves for $\lambda_c$ and $\theta$ obtained from more than 400 SM trajectories. A bi-exponential fit of the overall correlation curves (up to 50 sec as shown in pink shade) results two time constants, which are very similar to that observed earlier by Lu & Xie in 1997 [4]. An extremely long tail is observed in both the auto-correlation and cross-correlation curves for lag times exceeding 50 seconds. In the auto-correlation curves, this manifests as a negative correlation, likely resembling the well-known "cage effect." This effect possibly arises from the back-and-forth oscillatory motion (see Fig.S5) of molecules (or particles) trapped within a nanocavity (of cover slip silica surface and protecting PMMA polymer layer).

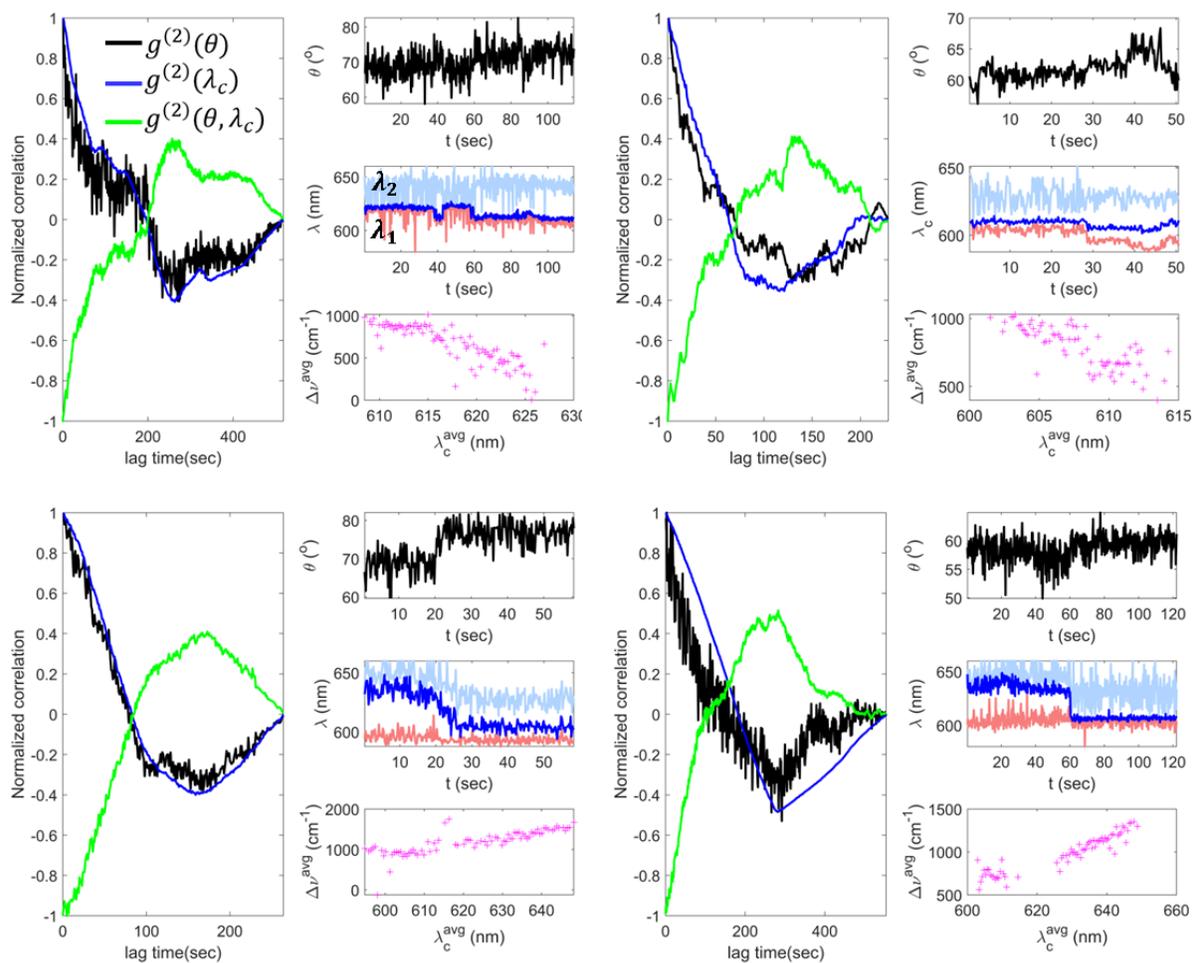

**Fig. S7.** Four representative single-molecule (SM) auto-correlation and cross-correlation curves for $\theta$ and $\lambda_c$ are presented, along with their respective time traces on the right side. Time traces for $\lambda_1$ and $\lambda_2$ are also included in the $\lambda_c$ plot for comparison. This figure illustrates two distinct subclasses of spectral jumps observed. In the top panel, two SM spectral trajectories are shown, where smaller spectral jumps are primarily driven by changes in $\lambda_1$, and the average energy of the associated vibronic bands ($\Delta\nu^{avg}$) decreases with an increasing spectral centroid $\lambda_c$. In the bottom panel, larger spectral jumps are exhibited by two other SM spectral trajectories, predominantly influenced by $\lambda_2$, with the mean energy of the associated vibronic bands ($\Delta\nu^{avg}$) increasing with $\lambda_c$. This behavioural difference is highlighted in Fig. S8.

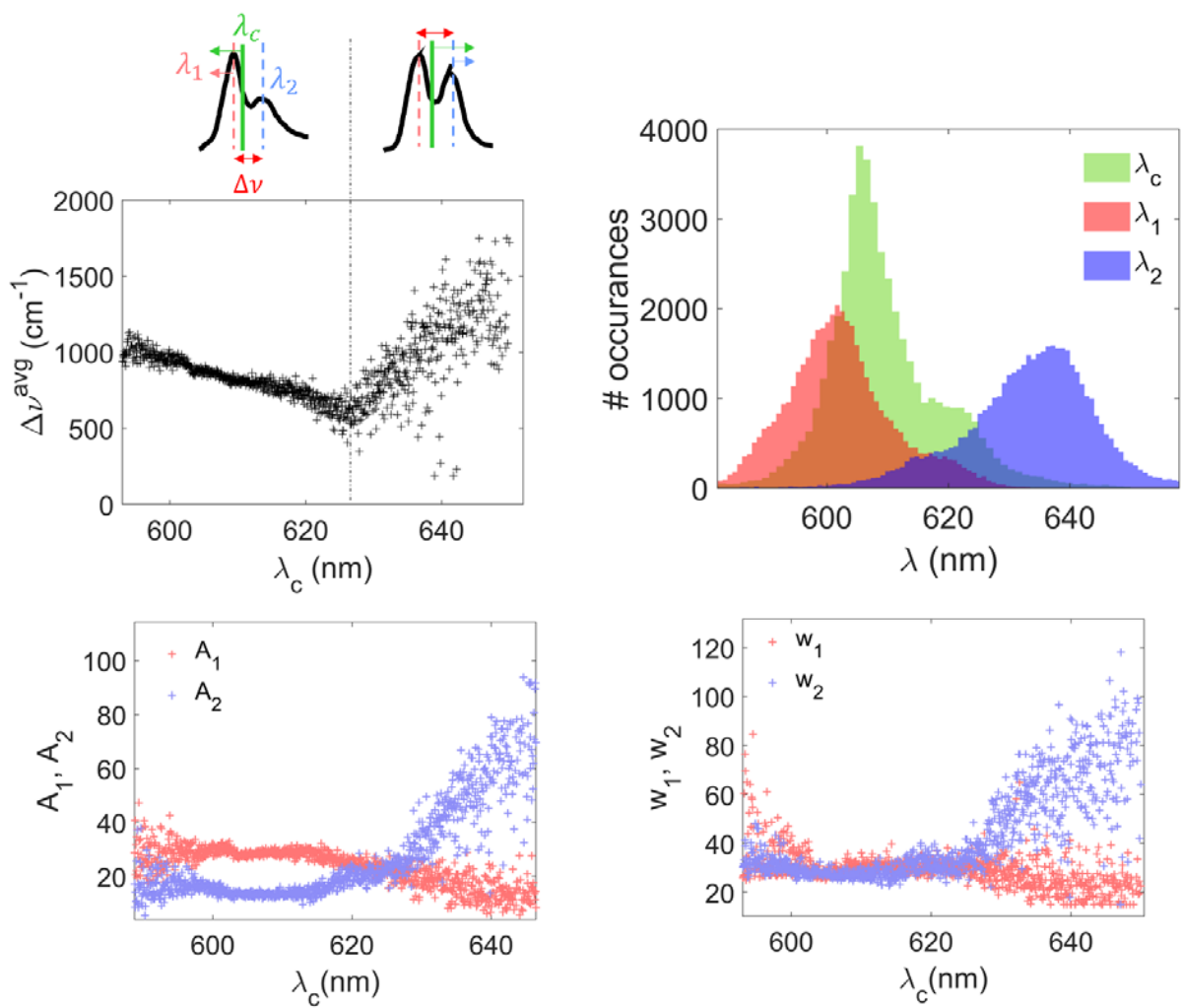

**Fig. S8.** The overall trend of the mean energy of associated vibronic bands with the spectral centroid is illustrated here, suggesting two distinct domains with respect to the spectral centroid. For $\lambda_c < 624$ nm, spectral fluctuations are predominantly contributed by the first component, $\lambda_1$, resulting in a decreasing nature of $\Delta\nu^{avg}$ with $\lambda_c$, as shown in the top left panel. For $\lambda_c > 624$ nm, the second component, $\lambda_2$, becomes the main contributor to spectral fluctuations, leading to an increasing trend of $\Delta\nu^{avg}$ with $\lambda_c$. The top right panel displays histograms of $\lambda_c$, $\lambda_1$ and $\lambda_2$, indicating two distinct contributions to $\lambda_c$, one around 610 nm and the other around 624 nm. The histograms of $\lambda_1$ and $\lambda_2$ clearly show that above 624 nm, $\lambda_2$ is the major contributor to $\lambda_c$, corroborating our observations (top left panel and Fig. S7). The histograms of integrated intensities, represented by the areas $A_1$ and $A_2$ of individual Gaussians along with their widths $w_1$ and $w_2$, shown in the bottom panel, further support these findings.

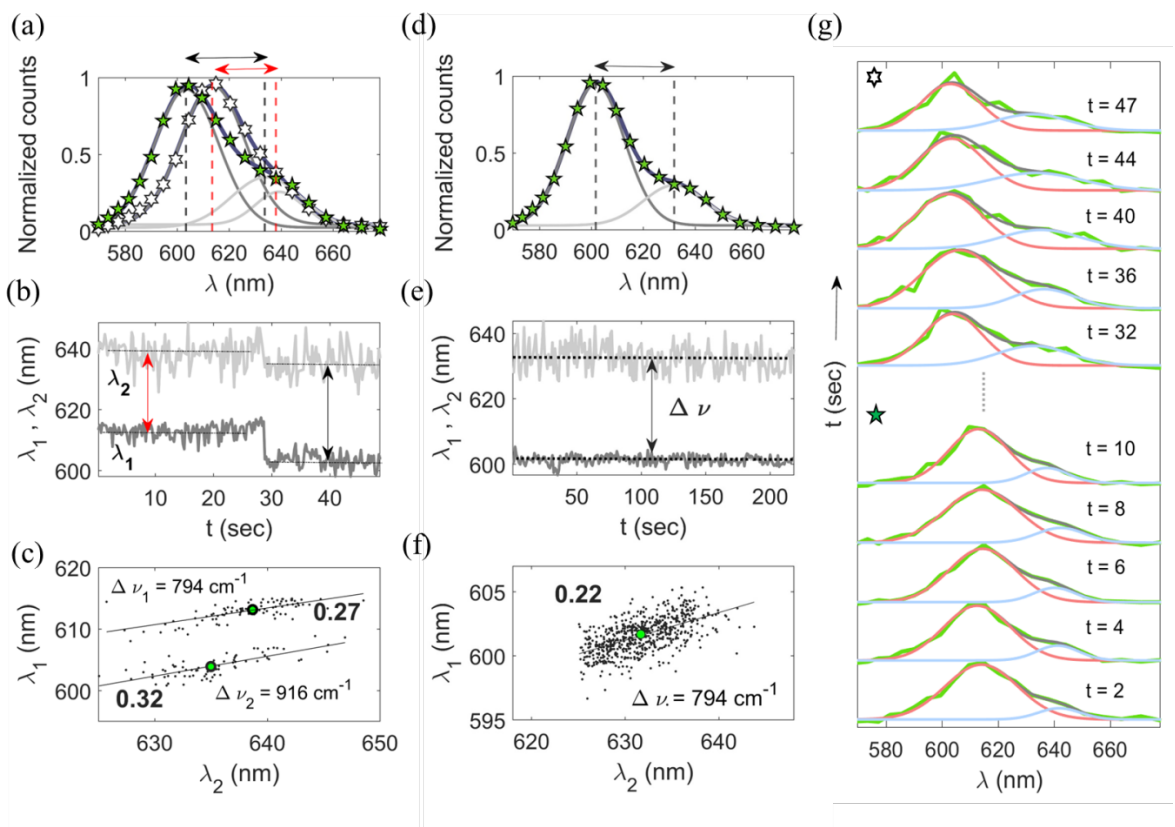

**Fig. S9.** (a) Time-averaged fitted spectra before and after the spectral jump at approximately t=30 seconds. (b,e) The time trace of these components, derived from the fitting of spectra in individual frames (shown in light colors). The black dotted line on top represents the fitted values of these components averaged over the respective time range. (c,f) 2D plots for vibronic band maxima of the corresponding random telegraphs. (d) Same as (a) except there is no spectral jump during its tenure. (g) Few of the fitted spectra before and after the jump at $t = 30\ sec$ for the same molecule shown in Fig. S9a, where we have represented fitted mean spectra for $t < 30\ s$ and $t > 30\ s$.

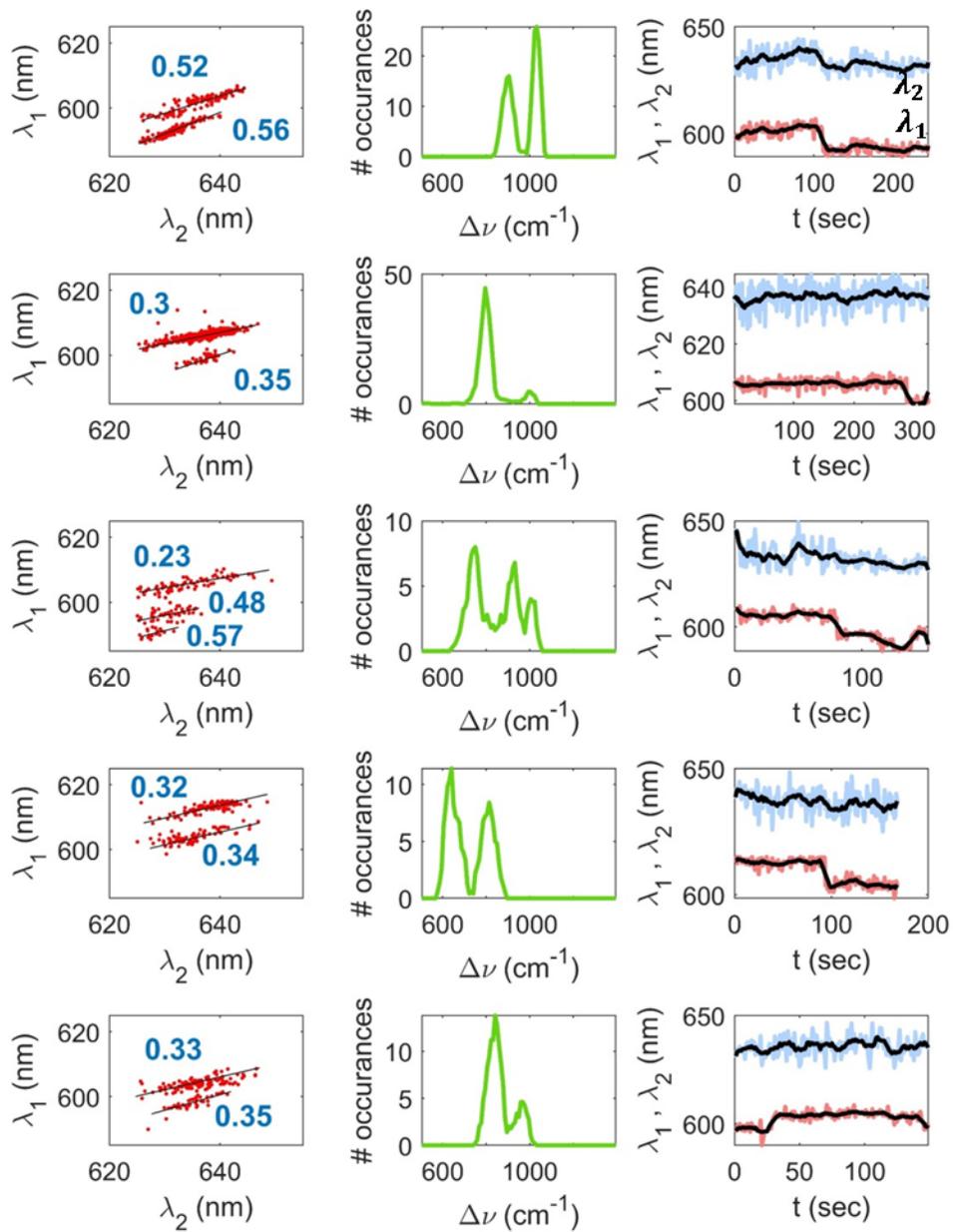

**Fig. S10.** 2D plots for vibronic band maxima's, corresponding vibrational spectrum and $\lambda_1, \lambda_2$ trajectories for five representative single molecules, showing spectral jump between various vibronic modes.

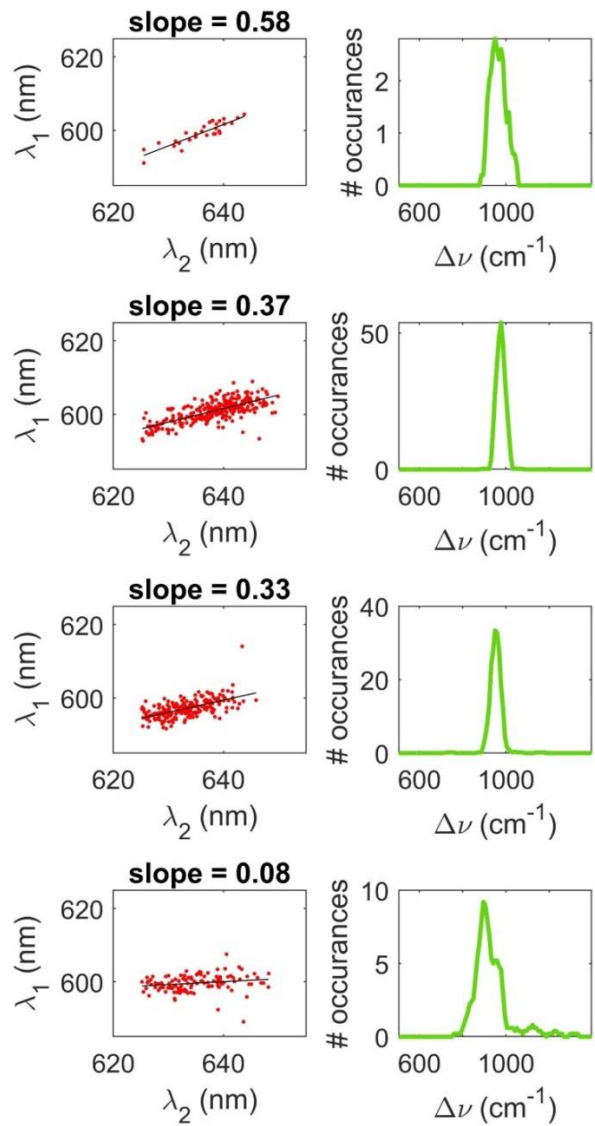

**Fig. S11.** 2D plots for vibronic band maxima's for few single molecules showing the same vibronic band around 917 cm$^{-1}$ but possesses different slope value, attributed to the influence of different local environment experienced by the emitter.

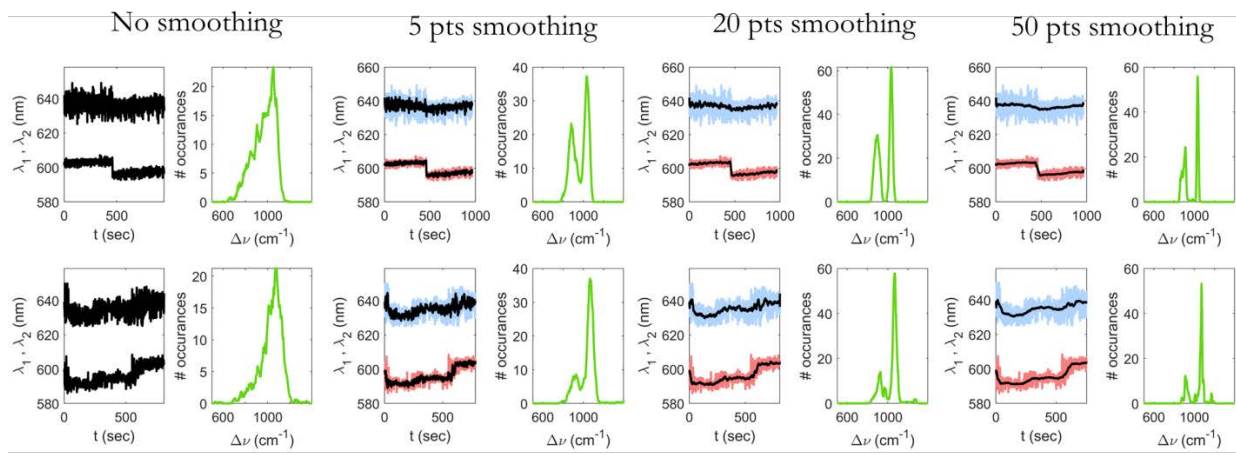

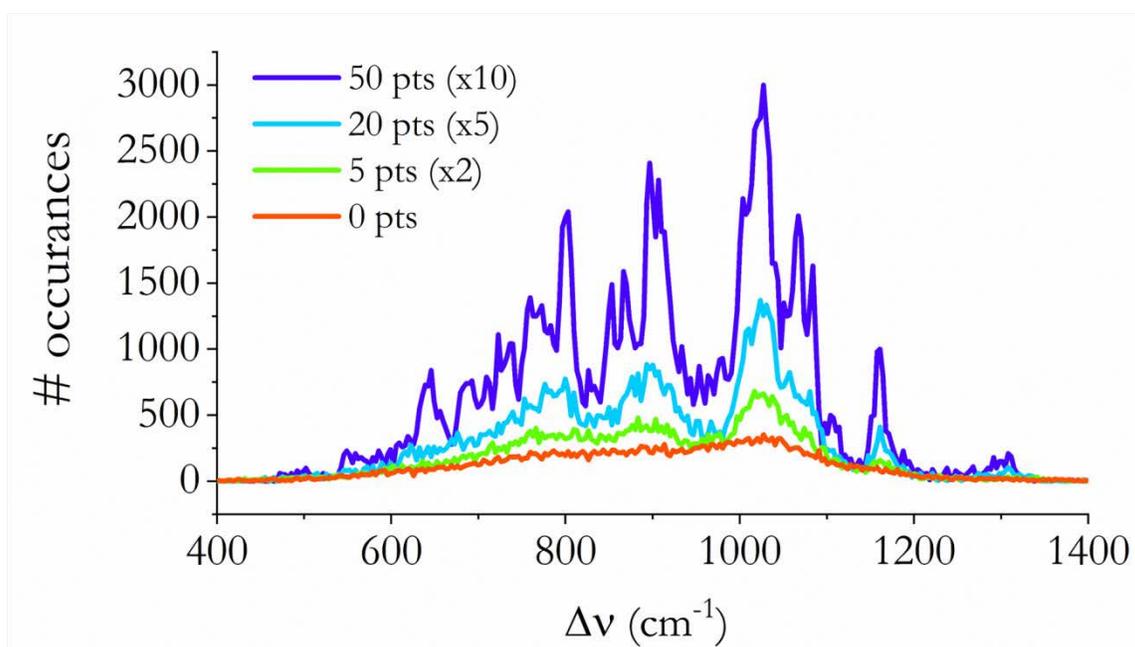

**Fig. S12.** Increasing level of smoothing (from left to right) using median filter increases the sharpness of the vibrational bands in the spectra, shown in top panel. Composite spectra combining contributions from the selected trajectories are depicted in the lower panel with similar smoothing chronology.

**Table S1:** Reported [5-8] and observed vibrational modes of Rhodamine 6G.

| Observed (cm⁻¹) | Reported (NR) / cm⁻¹ | | Reported (RR) / cm⁻¹ | | Description |
|---|---|---|---|---|---|
| | Experimental | Theoretical | Experimental | Theoretical | |
| 376 | 375 | | | | Torsional and bending motions [8] |
| 642 | 613 | 616 | 611 | 610 | ip XRD/op XRD [5] |
| 796 | 775 | 748, 771 | 775 | 765 | op C-H bend/ip XRD [5] |
| 870 | 903 | 895 | | | stretching modes of aromatic benzene rings [8] |
| 913 | 923 | | 934 | 917 | |
| 1033 | 1022 | 1033 | | | |
| 1074 | 1084 | 1091 | | | |
| 1102 | | 1115 | | | |
| 1178 | 1184 | 1185 | 1178 | 1185 | ip XRD, C-H bend, N-H bend [5] |
| 1308 | 1312 | 1346 | 1310 | 1297 | torsional and bending motions [8] |

*NR: Normal Raman; RR: Resonance Raman; ip: in-plane; op: out-of-plane; XRD: xanthene ring distortion*

## 5. Multi-dimensional imaging modality

As a proof of principle demonstration, we recorded a short 50 frames stack of spin coated Rh6G molecules for SM spectral and vibrational imaging along with their 3D dipole orientation and nanoscopic localization, shown in Fig. S13. The possibility of vibrational imaging and multiplexing is amply evident from Fig. S13b, respectively, together with spectral (Fig. S13f) and orientation imaging (Fig. S13c-e). Such multi-dimensional imaging modality is appropriate to explore nanoscopic structure and molecular interactions even in a highly heterogeneous matrix. Further investigations in these directions are ongoing.

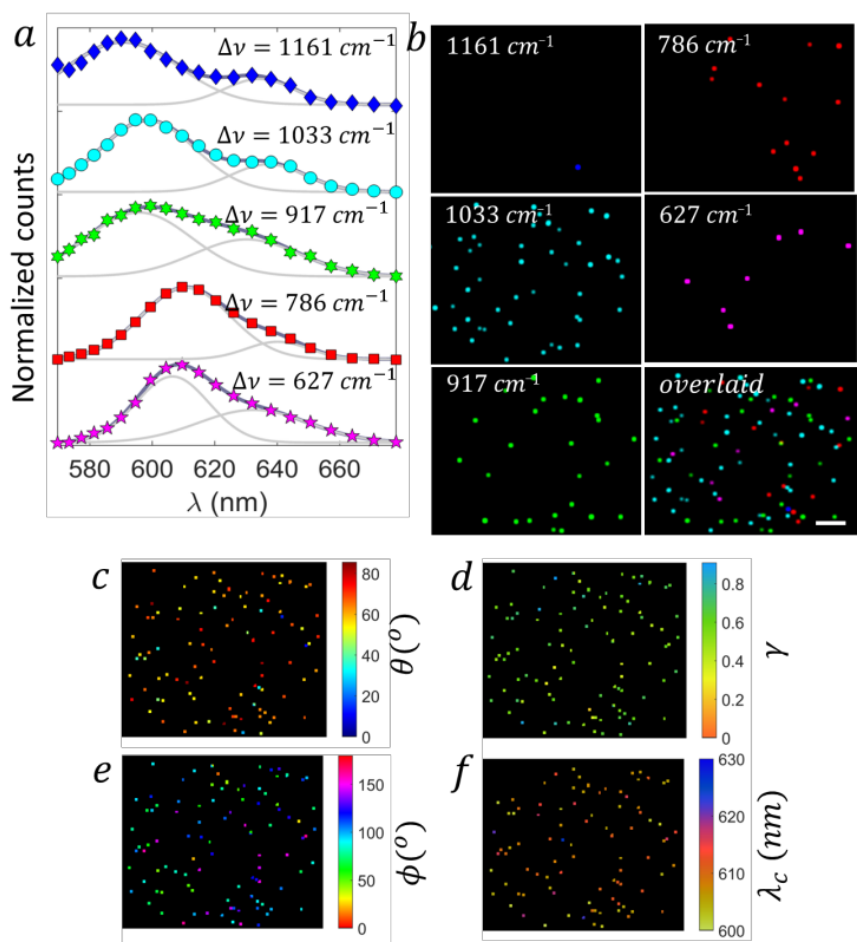

**Fig. S13.** (a) Representative double Gaussian fit of selected emission spectra for five color coded vibrational modes (blue, cyan, green, red & pink for 1161, 1033, 917, 786 & 627 cm 1, respectively), with their individual and overlaid images (b). For representation 0.1 pixel Gaussian blur is used. (c-f) Orientation and spectral images for the same region. Scale bar 2 μm.

Contrary to highly sophisticated and ultrafast laser-intensive vibrational techniques based on double resonance [9] to explore rich chemical information with optical contrast of single molecules, like stimulated Raman excited fluorescence (SREF) microscopy [10-14], fluorescence-encoded infrared (FEIR) spectroscopy [15-17], bond-selective fluorescence-detected infrared-excited (BonFIRE) spectro-microscopy [18], mid-infrared spectroscopy through vibrationally assisted luminescence [19], etc. the discussed SR-SMOLM [1] has a simple optical design with CW laser excitation and easy to adopt in any super-resolution microscopy setup based on single molecules localization. The latter however lacks temporal resolution achievable with ultrafast techniques. However, employing the principal of localization for bright emitters, SR-SMOLM inherently provides a superior lateral localization precession (~ 10 nm) for the extracted vibrational bands suitable for nanoscopy than reported with the above intricate techniques, e.g. BonFIRE [18] has ~600 nm lateral localization precession. Volumetric vibrational imaging (VISTA) following stimulated Raman scattering of endogenous proteins under isotropic expansion has demonstrated resolution down to 78 nm [20], but lacks multi-dimensional functional imaging capability to interrogate any molecular interactions. Moreover, photodamage with low power CW excitation in SR-SMOLM is expected to be negligible compared to high power femtosecond laser excitations. For only super-resolution vibrational imaging we can adopt to spectrally resolved localization microscopy [1, 21-31] with appropriate photon budgets, which helps in faster image acquisition, relevant with sensitive samples like live cell. A factor of convenience here is that majority of probes used for nanoscopy exhibit vibronic bands, other than isotopically labeled designer probes. Through suitable selection of dyes and appropriate excitation wavelength it is possible to select the accessible vibrational modes. Further investigations on all these aspects and improvements are ongoing.

**References:**

[1]     A. Sarkar, J. B. Mitra, V. K. Sharma, V. Namboodiri, and M. Kumbhakar, Spectrally Resolved Single Molecule Orientation Imaging Reveals Direct Correspondence between Polarity and Order Experienced by Nile Red in Supported Lipid Bilayer Membrane, bioRxiv 2024.06.21.600028; doi: https://doi.org/10.1101/2024.06.21.600028 (2024).

[2]     A. Sarkar, V. Namboodiri, and M. Kumbhakar, Single-Molecule Orientation Imaging Reveals Two Distinct Binding Configurations on Amyloid Fibrils, J. Phys. Chem. Lett. **14**, 4990 (2023).

[3]     C. N. Hulleman, R. Ø. Thorsen, E. Kim, C. Dekker, S. Stallinga, and B. Rieger, Simultaneous Orientation and 3D Localization Microscopy with a Vortex Point Spread Function, Nat. Comm. **12**, 5934 (2021).

[4]     H. P. Lu, and X. S. Xie, Single-molecule Spectral Fluctuations at Room Temperature, Nature **385** 143 (1997).

[5]     L. Jensen, and G. C. Schatz, Resonance Raman Scattering of Rhodamine 6G as Calculated Using Time-Dependent Density Functional Theory, J. Phys. Chem. A **110** 5973 (2006).

[6]     J. Guthmuller, and B. t. Champagne, Resonance Raman Scattering of Rhodamine 6G as Calculated by Time-Dependent Density Functional Theory: Vibronic and Solvent Effects, J. Phys. Chem. A **112**, 3215 (2008).

[7]     H. Watanabe, N. Hayazawa, Y. Inouye, and S. Kawata, DFT Vibrational Calculations of Rhodamine 6G Adsorbed on Silver: Analysis of Tip-Enhanced Raman Spectroscopy, J. Phys. Chem. B **109**, 5012 (2005).

[8]     S. Shim, C. M. Stuart, and R. A. Mathies, Resonance Raman Cross-Sections and Vibronic Analysis of Rhodamine 6G from Broadband Stimulated Raman Spectroscopy, ChemPhysChem **9**, 697 (2008).

[9]     A. Laubereau, A. Seilmeler, and W. Kaiser, A New Technique to Measure Ultrashort Vibrational Relaxation Times in Liquid Systems, Chem. Phys. Lett. **36**, 232 (1975).

[10]    Z. C. Lu Wei1, L. Shi, R. Long, A. V. Anzalone, L. Zhang, F. Hu, R. Yuste, V. W. Cornish, and W. Min, Super-multiplex Vibrational Imaging, Nature **544**, 465 (2017).

[11]    H. Xiong, and W. Min, Combining the Best of Two Worlds: Stimulated Raman Excited Fluorescence, J. Chem. Phys. **153**, 210901 (2020).

[12]    H. Xiong, L. Shi, L. Wei, Y. Shen, R. Long, Z. Zhao, and W. Min, Stimulated Raman Excited Fluorescence Spectroscopy and Imaging, Nature Photonics **13**, 412 (2019 ).

[13]    H. Xiong, N. Qian, Y. Miao, Z. Zhao, and W. Min, Stimulated Raman Excited Fluorescence Spectroscopy of Visible Dyes, J. Phys. Chem. Lett. **10**, 3563−3570 (2019).

[14]   H. Xiong, N. Qian, Y. Miao, Z. Zhao, C. Chen, and W. Min, Super-resolution Vibrational Microscopy by Stimulated Raman Excited Fluorescence, Light: Science & Applications **10**, 87 (2021).

[15]   L. Whaley-Mayda, A. Guha, S. B. Penwell, and A. Tokmakoff, Fluorescence-Encoded Infrared Vibrational Spectroscopy with Single-Molecule Sensitivity, J. Am. Chem. Soc. **143**, 3060−3064 (2021).

[16]   L. Whaley-Mayda, S. B. Penwell, and A. Tokmakoff, Fluorescence-Encoded Infrared Spectroscopy: Ultrafast Vibrational Spectroscopy on Small Ensembles of Molecules in Solution, J. Phys. Chem. Lett. **10**, 1967−1972 (2019).

[17]   A. Guha, L. Whaley-Mayda, S. Y. Lee, and A. Tokmakoff, Molecular Factors Determining Brightness in Fluorescence Encoded Infrared Vibrational Spectroscopy, J. Chem. Phys. **160**, 104202 (2024).

[18]   H. Wang, D. Lee, Y. Cao, X. Bi, J. Du, K. Miao, and L. Wei, Bond-selective Fluorescence Imaging with Single-molecule Sensitivity, Nat. Photonics **17** 846 (2023 ).

[19]   R. Chikkaraddy, R. Arul, L. A. Jakob, and J. J. Baumberg, Single-molecule Mid-infrared Spectroscopy and Detection Through Vibrationally Assisted Luminescence, Nat. Photon. **17** 865 (2023 ).

[20]   C. Qian, K. Miao, L.-E Lin, X. Chen, J. Du, and L. Wei, Super-resolution Label-free Volumetric Vibrational Imaging, Nat. Comm. **12**, 3648 (2021).

[21]   G.-h. Kim, J. Chung, H. Park, and D. Kim, Single-Molecule Sensing by Grating-based Spectrally Resolved Super-Resolution Microscopy, Bull. Korean Chem. Soc. **42**, 270 (2021).

[22]   D. Kim, Z. Zhang, and K. Xu, Spectrally Resolved Super-Resolution Microscopy Unveils Multipath Reaction Pathways of Single Spiropyran Molecules, J. Am. Chem. Soc. **139**, 9447 (2017).

[23]   C. Butler, G. E. Saraceno, A. Kechkar, N. Bénac, V. Studer, J. P. Dupuis, L. Groc, R. Galland, and J.-B. Sibarita, Multi-Dimensional Spectral Single Molecule Localization Microscopy, Front. Bioinform. **2**, 813494 (2022).

[24]   B. Brenner, C. Sun, F. M. Raymo, and H. F. Zhang, Spectroscopic Single‑molecule Localization Microscopy: Applications and Prospective, Nano Convergence **10**, 14 (2023).

[25]   B. Dong, L. Almassalha, B. E. Urban, T.-Q. Nguyen, S. Khuon, T.-L. Chew, V. Backman, C. Sun, and H. F. Zhang, Super-resolution Spectroscopic Microscopy via Photon Localization, Nat. Comm. **7**, 12290 (2016).

[26]   Z. Zhang, S. J. Kenny, M. Hauser, W. Li, and K. Xu, Ultrahigh-throughput Single-molecule Spectroscopy and Spectrally Resolved Super-resolution Microscopy, Nat. Methods **12**, 935 (2015).

[27]   Y. Zhang, Y. Zhang, K.-H. Song, W. Lin, C. Sun, G. C. Schatz, and H. F. Zhang, Investigating Single-Molecule Fluorescence Spectral Heterogeneity of Rhodamines Using High-Throughput Single-Molecule Spectroscopy, J. Phys. Chem. Lett. **12**, 3914−3921 (2021).

[28]   P. M. Lundquist, C. F. Zhong, P. Zhao, A. B. Tomaney, P. S. Peluso, J. Dixon, B. Bettman, Y. Lacroix, D. P. Kwo, E. McCullough, M. Maxham, K. Hester, P. McNitt, D. M. Grey, C. Henriquez, M. Foquet, S. W. Turner, and D. Zaccarin, Parallel Confocal Detection of Single Molecules in Real Time, Opt. Lett. **33**, 1026 (2008).

[29]   J. Eid, A. Fehr, J. Gray, K. Luong, J. Lyle, G. Otto, P. Peluso, D. Rank, P. Baybayan, B. Bettman, A. Bibillo, K. Bjornson, B. Chaudhuri, F. Christians, R. Cicero, S. Clark, R. Dalal, A. deWinter, J. Dixon, M. Foquet, A. Gaertner, P. Hardenbol, C. Heiner, K. Hester, D. Holden, G. Kearns, X. Kong, R. Kuse, Y. Lacroix, S. Lin, P. Lundquist, C. Ma, P. Marks, M. Maxham, D. Murphy, I. Park, T. Pham, M. Phillips, J. Roy, R. Sebra, G. Shen, J. Sorenson, A. Tomaney, K. Travers, M. Trulson, J. Vieceli, J. Wegener, D. Wu, A. Yang, D. Zaccarin, P. Zhao, F. Zhong, J. Korlach, and S. Turner, Real-Time DNA Sequencing from Single Polymerase Molecules, Science **323** 133 (2009).

[30]   J. Jeffet, A. Ionescu, Y. Michaeli, D. Torchinsky, E. Perlson, T. D. Craggs, and Y. Ebenstein, Multimodal Single-molecule Microscopy with Continuously Controlled Spectral Resolution, Biophysical Reports **1**, 100013 (2021).

[31]   M. N. Bongiovanni, J. Godet, M. H. Horrocks, L. Tosatto, Alexander R. Carr, D. C. Wirthensohn, R. T. Ranasinghe, J.-E. Lee, A. Ponjavic, J. V. Fritz, C. M. Dobson, D. Klenerman, and S. F. Lee, Multi-dimensional Super-resolution Imaging Enables Surface Hydrophobicity Mapping, Nat. Comm. **7**, 13544 (2016).